\documentclass{article}

\usepackage{microtype}
\usepackage{graphicx}
\usepackage{subcaption}
\usepackage{booktabs} 
\usepackage{natbib}
\usepackage{xcolor}
\usepackage{commath}
\usepackage{xcolor}
\usepackage{tabularx,lipsum}
\usepackage{booktabs}
\usepackage{makecell}
\usepackage{hyperref}

\usepackage[accepted]{icml2020}

\icmltitlerunning{A Deep Learning Approach to Operational Flare Forecasting}
\newcolumntype{R}{>{\raggedright\arraybackslash}X}
\begin{document}

\twocolumn[
\icmltitle{A Deep Learning Approach to Operational Flare Forecasting} 


\begin{icmlauthorlist}
\icmlauthor{Yasser Abduallah}{goo}
\icmlauthor{Jason T. L. Wang}{too}

\end{icmlauthorlist}

\icmlaffiliation{goo}{Department of Computer Science, New Jersey Institute of Technology, Newark, NJ 07102, USA}
\icmlaffiliation{too}{College of Computing, New Jersey Institute of Technology, Newark, NJ 07102, USA}

\icmlcorrespondingauthor{Jason T. L. Wang}{wangj@njit.edu}

\icmlkeywords{Machine Learning}

\vskip 0.3in
]


\printAffiliationsAndNotice{}  

\begin{abstract}
Solar flares are explosions on the Sun.
They happen when energy stored in magnetic fields
around solar active regions (ARs) is suddenly released.
In this paper, we present a transformer-based framework, named SolarFlareNet,
for predicting whether an AR would
produce a $\gamma$-class flare within the next 24 to 72 hours.
We consider three $\gamma$ classes, namely the $\geq$M5.0 class, the $\geq$M class and the $\geq$C class, and build three transformers separately, 
each corresponding to a $\gamma$ class. 
Each transformer is used to
make predictions of its corresponding $\gamma$-class flares.
The crux of our approach is to model data samples in an
AR as time series and to use transformers
to capture the temporal dynamics of the data samples.
Each data sample consists of magnetic parameters taken from
Space-weather HMI Active Region Patches (SHARP) and related data products. 
We survey flare events that occurred from May 2010 to December 2022
using the Geostationary Operational Environmental Satellite X-ray flare catalogs provided by
the National Centers for Environmental Information (NCEI),
and build a database of flares
with identified ARs in the NCEI flare catalogs.
This flare database is used to construct labels of the data samples suitable for machine learning.
We further extend the deterministic approach to a calibration-based probabilistic forecasting method.
The SolarFlareNet system is fully operational and is capable of making near real-time predictions of
solar flares on the Web.
\end{abstract}

\section{Introduction} \label{sec:intro}
Solar flares are sudden explosions of energy that occur on the Sun's surface.
They often occur in solar active regions (ARs), caused by
strong magnetic fields typically associated with sunspot areas.
Solar flares are categorized into five classes
A, B, C, M, and X, with A-class flares having the lowest intensity and X-class flares
having the highest intensity.
Major flares are usually accompanied by 
coronal mass ejections and solar energetic particles 
\cite{CME2008SoPh..248..471Q,BobraCME2016ApJ...821..127B,Liu..Wang..Solar..2017ApJ...843..104L,Liu_2020CMEPrediction,AbduallahSEP2022,SEP2022cosp...44.1151M}. 
These eruptive events can have significant and harmful effects on or near Earth, damaging technologies, power grids, space stations, and human life
\cite{Daglis:2004:EffectSpace,Liu_2019FlarePrediction,FlareIndex2022ApJS..263...28Z,FlareForecasting2023ChAA..47..108H}. 
Therefore, providing accurate and early forecasts of solar flares
is crucial for disaster risk management, risk mitigation, and preparedness.

Although a lot of effort has been devoted to flare prediction
\cite{ImprovingFlarePredictionHuang,Panos_2020,FlareLikelyHoodGeorgoulis,Tang_2021},
developing accurate, operational near-real-time flare forecasting systems remains a
challenge.
In the past, researchers designed
statistical models for the prediction of flares based on the physical properties of active regions \cite{Gallagher2002SoPh..209..171G,LekaBarnes2007ApJ...656.1173L,Manson2010ApJ...723..634M}.
With the availability of large amounts of flare-related data
\cite{FlareLikelyHoodGeorgoulis}, researchers started using machine learning methods for flare forecasting \cite{Bobora:2015ApJ...798..135B,Liu..Wang..Solar..2017ApJ...843..104L,MLaaS_RAA_Abduallah_2021}.
More recently, deep learning, which is a subfield of machine learning, has
emerged and showed promising results in predicting solar eruptions,
including solar flares \cite{LiuXu2022ApJ...941...20L,Sun3DRNN2022ApJ...941....1S}. 

For example, Nishizuka et al. \cite{2018ApJ...858..113N_Nishizuka} developed deep neural networks to forecast M- and C-class flares that would occur within 24 hours using data
downloaded from the Solar Dynamics Observatory (SDO) \cite{PesnellSDO2015hchp} and the Geostationary Operational Environmental Satellite (GOES). 
Sun et al. \cite{Sun3DRNN2022ApJ...941....1S} employed three-dimensional (3D) 
convolutional neural networks (CNNs) to forecast $\geq$M-class and $\geq$C-class flares
using Space-weather HMI Active Region Patches 
(SHARP) \cite{Bobra:2014SoPh..289.3549B}
magnetograms downloaded from the Joint Science Operations Center (JSOC)
accessible at \url{http://jsoc.stanford.edu/}.
Li et al. \cite{LiFlareCNN2020ApJ...891...10L} also adopted a CNN model to forecast
$\geq$M-class and $\geq$C-class flares
using SHARP magnetograms where
the authors restructured the CNN layers in their neural network
with different filter sizes.
Deng et al. \cite{Deng2021ApJ...922..232D} developed a hybrid CNN model to predict solar flares
during the rising and declining phases of Solar Cycle 24. 

Some researchers adopted SHARP magnetic parameters in time series for flare prediction.
Static SHARP parameters quantitatively describe the properties of ARs,
especially their ability to produce flares,
at a given time. 
On the other hand, dynamic information, such as
the magnetic helicity injection rate, sunspot motions, shear flows, and magnetic
flux emergence/flux cancelation,
is more important for flare forecasting.  
Using time series of SHARP parameters allows a model to
capture the relationship between the evolution of magnetic fields of ARs and solar flares,
hence achieving more accurate flare predictions~\cite{Tian_2022_RelationBetweenSolarFieldsFlaresinARS, EvolutionOfARs}.
In an earlier study,
Yu et al. \cite{SequentialSupervisedLearningYuEtAl2009SoPh}
added the evolutionary information of ARs to
a predictive model for the prediction of short-term solar flares.
More recently,
Chen  et al. \cite{Chen2019LSTMFlare} designed a long short-term memory (LSTM) network to identify precursors of
solar flare events using time series of SHARP parameters.
LSTM is suitable for capturing the temporal dynamics of time series.
Liu  et al.  \cite{Liu_2019FlarePrediction} developed another LSTM network with a customized attention mechanism 
to direct the network to focus on important patterns 
in time series of SHARP parameters.
Sun et al. \cite{FlarePredictionCNNLSTMSun_2022} attempted to distinguish between ARs with strong flares ($\geq$M-class flares) and ARs with no flare at all.
The authors showed that combining LSTM and CNN can better solve
the ``strong versus quiet'' flare prediction problem,
with data from both Solar Cycle 23 and Cycle 24.
All of the aforementioned studies provided valuable models and algorithms in the field. 
However, the existing methods focused on short-term forecasts (usually within 24 hours).
Furthermore, the models were not used as operational systems.

In this paper, we propose a new deep learning approach to predicting solar flares using time series of SHARP parameters.
Our approach employs a transformer-based framework, named SolarFlareNet, which
predicts whether there would be a flare within 24 to 72 hours, where the flare could be a 
$\geq$M5.0-, $\geq$M-, or
$\geq$C-class flare.
We further extend SolarFlareNet to produce probabilistic forecasts of flares and
implement the probabilistic model into an
operational, near real-time flare forecasting system.
Experimental results demonstrate that SolarFlareNet
generally performs better than,
or is comparable to,
related flare prediction methods.

\section{Results}
\label{sec:solarflarenetresults}

\subsection{Deterministic Prediction Tasks} 
For any given active region (AR) and time point $t$, 
we predict whether there would be a $\gamma$-class flare 
within the next 24 hours (48 hours, 72 hours, respectively) of $t$
where $\gamma$ can be $\geq$M5.0, $\geq$M, or $\geq$C.
A $\geq$M5.0-class flare
means the GOES X-ray flux value of the flare is above
$5 \times 10^{-5}\mbox{Wm}^{-2}$. A $\geq$M-class flare refers to an X- or M-class flare.
A $\geq$C-class flare refers to an X-class, M-class, or C-class flare.
We focus on these three classes of flares due to their importance in space weather
\cite{Bobora:2015ApJ...798..135B,2018SoPh..293...48J,2018ApJ...858..113N_Nishizuka,Liu_2019FlarePrediction}.
We developed three transformer models
to tackle the three
prediction tasks individually and separately. 
Notice that we did not consider $\gamma$ to be $\geq$X
due to the lack of samples for X-class flares.
Instead, we use $\geq$M5.0 
as the most significant class, 
which contains sufficient samples. 

\subsection{Comparison with Previous Methods}
\label{sec:solarflarenetcomparativestudy}
We conducted a series of
experiments to compare the proposed SolarFlareNet framework
with closely related methods.
All these methods perform binary classifications/predictions as defined above.
We adopt several performance metrics. Formally, given an AR and a data sample $x_{t}$ observed at time point $t$, we define $x_{t}$ to be a true positive (TP) if 
the $\ge$M5.0 ($\ge$M, $\ge$C, respectively) model
predicts that $x_{t}$ is positive, 
i.e., the AR will produce a 
$\ge$M5.0- ($\ge$M-, $\ge$C-, respectively) class
flare within the next 24 hours
of $t$, and
$x_{t}$ is indeed positive.
We define $x_{t}$ as a false positive (FP) if 
the $\ge$M5.0 ($\ge$M, $\ge$C, respectively) model
predicts that $x_{t}$ is positive
while $x_{t}$ is actually negative,
i.e., the AR will not produce a 
$\ge$M5.0- ($\ge$M-, $\ge$C-, respectively) class
flare within the next 24 hours of $t$.
We say $x_{t}$ is a true negative (TN) if 
the $\ge$M5.0 ($\ge$M, $\ge$C, respectively) model
predicts $x_{t}$ to be negative
and $x_{t}$ is indeed negative;
$x_{t}$ is a false negative (FN) if 
the $\ge$M5.0 ($\ge$M, $\ge$C, respectively) model
predicts $x_{t}$ to be negative while $x_{t}$ is actually positive.
We also use TP (FP, TN, and FN, respectively) to represent the total number of
true positives (false positives, true negatives, and false negatives, respectively).
The TP, FP, TN, and FN for the 48-hour and 72-hour ahead predictions are defined similarly. 
The performance metrics are calculated as follows:

\begin{equation}
	\text{Recall} = \frac{ \mathrm{TP}}{\mathrm{TP + FN}}
\end{equation}
	\begin{equation}
	\text{Precision} = \frac{ \mathrm{TP}}{\mathrm {TP + FP}}
\end{equation}
    \begin{equation}
		\text{Accuracy (ACC)} =  \frac{ \mathrm{TP + TN}}{\mathrm{TP+FP + TN + FN}}
	\end{equation}
	\begin{equation}
		\text{Balanced ACC (BACC)} = 
\frac{
  \left( \frac{ \mathrm{TP}}{\mathrm{TP+FN}} + \frac{ \mathrm{TN}}{\mathrm{TN+FP}}\right)}
  {2}
	\end{equation}
	\begin{equation}
		\text{True Skill Statistics (TSS)} = \frac{ \mathrm{TP}}{\mathrm{TP+FN}} - \frac{\mathrm{FP}}{\mathrm{FP+TN}}
	\end{equation}

Table \ref{tab:solarflarenetcomparativestudy2019sun3d} 
compares SolarFlareNet with related methods for 24-hour ahead 
flare predictions.
The performance metric values of SolarFlareNet are 
mean values obtained from
10-fold cross-validation
\cite{Liu_2019FlarePrediction}.
The metric values of the highest performance models
in the related studies
are taken directly from those studies and are
displayed in Table \ref{tab:solarflarenetcomparativestudy2019sun3d}.
The symbol `---' means that a method does not
produce the metric value for the corresponding prediction task.
The best metric values
are highlighted in boldface.
TSS is the primary metric used in the literature on flare prediction.
It can be seen from Table~\ref{tab:solarflarenetcomparativestudy2019sun3d} 
that SolarFlareNet outperforms the state-of-the-art methods
in terms of TSS except that Liu et al. \cite{Liu_2019FlarePrediction}
perform better than SolarFlareNet in predicting 
$\geq$M5.0 class flares.

\begin{table*}
\centering
\begin{tabular}{llccc}
\hline
Metric & Method & $\geq$M5.0 class  & $\geq$M class  & $\geq$C class  \\  
\hline
         Recall &   Huang  et al. \cite{HuangCNN2018ApJ...856....7H} 
         & --- & --- & --- \\
          & Li  et al. \cite{LiFlareCNN2020ApJ...891...10L} & --- & 0.817 & 0.889\\
         & Liu  et al. \cite{Liu_2019FlarePrediction} &\textbf{}\textbf{0.960}& 0.885 & 0.773\\
            & Sun  et al. \cite{Sun3DRNN2022ApJ...941....1S} & --- & \textbf{0.925} & 0.862\\
            & Wang  et al. \cite{Wang_2020} & --- & 0.730 & 0.621 \\
               & This work & 0.853 & 0.842 & \textbf{0.891} \\
        \hline
         Precision &  Huang  et al.  \cite{HuangCNN2018ApJ...856....7H} & --- & --- & --- \\
          & Li  et al. \cite{LiFlareCNN2020ApJ...891...10L} & --- & \textbf{0.889} & 0.906 \\
          & Liu  et al. \cite{Liu_2019FlarePrediction} & 0.048 & 0.222 & 0.541\\
                & Sun  et al. \cite{Sun3DRNN2022ApJ...941....1S} & ---   & 0.595 & 0.878\\
                &  Wang  et al. \cite{Wang_2020} & --- & 0.282 & 0.541 \\
               & This work & \textbf{0.977} & 0.848 & \textbf{0.949} \\
        \hline
         ACC  &   Huang  et al. \cite{HuangCNN2018ApJ...856....7H} & --- & --- & --- \\
          & Li  et al. \cite{LiFlareCNN2020ApJ...891...10L} & --- & 0.891 & 0.861 \\
          & Liu  et al.  \cite{Liu_2019FlarePrediction} &0.921 & 0.907& 0.826\\
                & Sun  et al. \cite{Sun3DRNN2022ApJ...941....1S} & --- & 0.904 & 0.879\\
                &  Wang  et al. \cite{Wang_2020} & --- &  \textbf{0.945}&  0.883\\
               & This work &\textbf{0.964}  & 0.928 & \textbf{0.915} \\         
         \hline
         BACC &   Huang  et al.    \cite{HuangCNN2018ApJ...856....7H} 
         & --- & --- & --- \\
          & Li  et al. \cite{LiFlareCNN2020ApJ...891...10L} & --- &
          --- & --- \\
          & Liu  et al. \cite{Liu_2019FlarePrediction} & \textbf{0.940} &0.896 & 0.806\\
                & Sun  et al. \cite{Sun3DRNN2022ApJ...941....1S}    
                &  ---     & ---  & ---  \\
            &   Wang  et al. \cite{Wang_2020} & --- & --- & --- \\
               & This work &  0.926 & \textbf{0.919} & \textbf{0.917} \\         
         \hline
         TSS  &  Huang  et al.     \cite{HuangCNN2018ApJ...856....7H} 
         & --- & 0.662 & 0.487 \\
          & Li  et al. \cite{LiFlareCNN2020ApJ...891...10L} & --- & 0.749 & 0.679\\
          & Liu  et al. \cite{Liu_2019FlarePrediction} &\textbf{0.881} &0.792 & 0.612\\
                & Sun  et al. \cite{Sun3DRNN2022ApJ...941....1S}   
                & ---   & 0.826 & 0.756\\
                & Wang  et al. \cite{Wang_2020} & --- & 0.681 & 0.553 \\
               & This work &0.818  & \textbf{0.839} & \textbf{0.835} \\ 
         \hline
\end{tabular}
\caption{Performance comparison between SolarFlareNet and related methods for 24-hour ahead flare predictions.}
\label{tab:solarflarenetcomparativestudy2019sun3d}
\end{table*}

Table \ref{tab:solarflarenet4872metrics} presents the
mean performance metric values with standard deviations
enclosed in parentheses
for the 48- and 72-hour forecasts made by SolarFlareNet.
None of the existing methods in Table \ref{tab:solarflarenetcomparativestudy2019sun3d}
provides predictions in 48 or 72 hours in advance and, therefore, they are not listed
in Table \ref{tab:solarflarenet4872metrics}.
Overall, SolarFlareNet performs well for the 48- and 72-hour forecasts. 
However, the metric values of the tool in Table \ref{tab:solarflarenet4872metrics} are lower than
those in Table \ref{tab:solarflarenetcomparativestudy2019sun3d}.
This is understandable due to the longer range of predictions in Table \ref{tab:solarflarenet4872metrics}.

\begin{table*}
\centering
\begin{tabular}{llccc}
\hline
Metric & Hour & $\geq$M5.0 class  & $\geq$M class  & $\geq$C class  \\  
\hline
         Recall 
                & 48 &0.739 (0.108) & 0.735 (0.089) & 0.722 (0.089) \\
                & 72 & 0.717 (0.100) & 0.708 (0.078) &  0.702 (0.089)\\
        \hline
         Precision 
                & 48 & 0.890 (0.210) & 0.823 (0.092) & 0.812 (0.072)\\
                & 72 &0.872 (0.045) &0.812 (0.089) & 0.809 (0.051) \\
        \hline
         ACC  
                & 48 & 0.923 (0.003) & 0.907 (0.007) & 0.896 (0.047)\\
                & 72& 0.906 (0.002) & 0.883 (0.005)& 0.863 (0.040) \\      
         \hline
         BACC 
                & 48 & 0.864 (0.054) & 0.857 (0.045) & 0.848 (0.040)\\
                & 72 & 0.856 (0.039) & 0.843 (0.048)& 0.834 (0.029)\\        
         \hline
         TSS 
                & 48 & 0.736 (0.112) &0.728 (0.090)  & 0.719 (0.079)\\
                & 72 & 0.729 (0.108) & 0.714 (0.095)& 0.709 (0.058)\\
         \hline
\end{tabular}
\caption{Performance metric values of SolarFlareNet for 48- and 72-hour ahead flare predictions.}
\label{tab:solarflarenet4872metrics}
\end{table*}

\subsection{Probabilistic Forecasting with Calibration}
\label{sec:solarflarenetprobabilisticprediction}
SolarFlareNet is essentially a probabilistic forecasting method, producing a probability between 0 and 1.
The method compares the probability with a predetermined threshold, which is set to 0.5.
Given an AR and a data sample $x_{t}$ at time point $t$,
if the predicted probability is greater than or equal to the threshold,
then the AR will produce a flare within the next 24 (48, 72, respectively) hours of $t$
(i.e., $x_{t}$ belongs to the positive class);
otherwise, the AR will not produce a flare within the next 24 (48, 72, respectively) hours of $t$
(i.e., $x_{t}$ belongs to the negative class).
We can turn SolarFlareNet into a probabilistic forecasting method by directly outputting
the predicted probability without comparing it with the threshold.
Under this circumstance, the output is interpreted as
a probabilistic estimate of how likely the AR will produce a flare within
the next 24 (48, 72, respectively) hours of $t$.
We employ a probability calibration technique with isotonic regression~\cite{Isotonic1967,IsotonicSager1982} to adjust the
predicted probability and avoid the mismatch between the distributions of the predicted and expected probabilistic values~\cite{AbduallahSEP2022}. 
We add a suffix ``-C'' to SolarFlareNet to
denote the network without calibration. 

To evaluate the performance of a probabilistic forecasting method, we use
the Brier score (BS) and
Brier skill score (BSS), defined as follows
\cite{BSSWilks2010,Liu_2020CMEPrediction,AbduallahSEP2022}: 
\begin{equation}
\text{BS} = \frac{1}{N}\sum_{i=1}^{N}(y_i - \hat{y}_i)^2
\end{equation}
\begin{equation}
\text{BSS} = 1 - \frac{BS}{\frac{1}{N}\sum_{i=1}^{N}(y_i - \bar{y})^2}
\end{equation}
where $N$ is the number of data samples in
a test set; $y_i$ denotes the observed probability and $\hat{y}_i$
denotes the predicted probability of the $i$th test data sample, respectively;
$\bar{y} = \frac{1}{N}\sum_{i=1}^{N} y_i$ denotes the mean of all the observed probabilities.
BS values range from 0 to 1, with 0 being a perfect score.
BSS values range from
$-\infty$ to 1, with 1 being a perfect score.

Table \ref{tab:probcalib}
compares SolarFlareNet, used as a probabilistic
forecasting method, with a
closely related method \cite{Liu_2019FlarePrediction}. 
The BS and BSS values in the table are mean values
(with standard deviations enclosed in parentheses)
obtained from 10-fold cross-validation.
The metric values for the existing method are taken directly from the related work \cite{Liu_2019FlarePrediction}. 
The best BS and BSS values are highlighted in bold.
Notice that the existing method did not make 48-hour or 72-hour forecasts in advance.
Table \ref{tab:probcalib} shows that
there is no definitive conclusion regarding the
relative performance of SolarFlareNet and the existing method.
The existing method is better in terms of BS, while
SolarFlareNet is better in terms of BSS.
On the other hand, the calibrated version of a model is better than the model without calibration. 
Notice also that the results of the 48-hour and 72-hour forecasts are worse than those
of the 24-hour forecasts. 
This is understandable since
the longer the prediction window, the worse the performance a model achieves due to data deviation over time.

\begin{table*}
\centering
\begin{tabular}{cllccc}
\hline
       Hour   & Metric & Method & $\geq$M5.0 class  & $\geq$M class  & $\geq$C class  \\  
\hline
24   &        &                   &                &      &            \\
          & BS     &  Liu  et al. \cite{Liu_2019FlarePrediction}  & \textbf{0.090} (0.011) & \textbf{0.090} (0.009) & \textbf{0.133} (0.007) \\
          &        & SolarFlareNet     & 0.226 (0.024) & 0.244 (0.013)  & 0.285 (0.034) \\  
          &        & SolarFlareNet-C   & 0.263 (0.024)          &  0.281  (0.050)         &  0.313 (0.033)   \\
         \hline
        & BSS      & Liu  et al. \cite{Liu_2019FlarePrediction}  & $-21.576$ (2.956) & $-2.241$ (0.319) &  0.152 (0.047)  \\
        &          & SolarFlareNet & \textbf{0.584} (0.022) & \textbf{0.521 (0.042)} 
        &\textbf{0.409} (0.062) \\               
        &          & SolarFlareNet-C       & 0.504 (0.026) & 0.491 (0.031)  & 0.349 (0.055)\\
         \hline
48  &         &                   &                &          &    \\
        &  BS   & Liu  et al. \cite{Liu_2019FlarePrediction} & --- & --- & ---         \\
        &       & SolarFlareNet & \textbf{0.272} (0.091) & \textbf{0.312} (0.101) & \textbf{0.361} (0.091)  \\        
        &       & SolarFlareNet-C & 0.315 (0.049)       & 0.336 (0.033)          & 0.378 (0.104)  \\
         \hline
        & BSS  & Liu  et al. \cite{Liu_2019FlarePrediction} & --- & --- &  ---\\
        &      & SolarFlareNet & \textbf{0.569} (0.045) & \textbf{0.524} (0.021)   & \textbf{0.502} (0.033) \\        
        &      & SolarFlareNet-C & 0.457 (0.062)     & 0.424 (0.091) & 0.411 (0.056)\\
         \hline
72 &         &                   &         &           &            \\
        &  BS   & Liu  et al.  \cite{Liu_2019FlarePrediction} & --- & --- & ---\\
        &       & SolarFlareNet           & \textbf{0.313} (0.062)  & \textbf{0.327} (0.063)  & \textbf{0.344} (0.049) \\
        &       & SolarFlareNet-C       & 0.329 (0.094) & 0.369 (0.088) & 0.376 (0.102)  \\
         \hline
        & BSS  & Liu  et al. \cite{Liu_2019FlarePrediction} & --- & --- & --- \\
        &      & SolarFlareNet                             & \textbf{0.549} (0.067) & \textbf{0.524} (0.089)  & \textbf{0.501} (0.093)  \\        
        &      & SolarFlareNet-C                         & 0.514 (0.077) & 0.469 (0.095)  & 0.447 (0.059)  \\
         \hline
\end{tabular}
\caption{Performance comparison between SolarFlareNet and an existing method for probabilistic flare predictions 
(24 to 72 hours in advance).}
\label{tab:probcalib}
\end{table*}

\subsection{The SolarFlareNet System}

\begin{figure*}[ht!]
	\centering
         \includegraphics[width=0.9\linewidth]{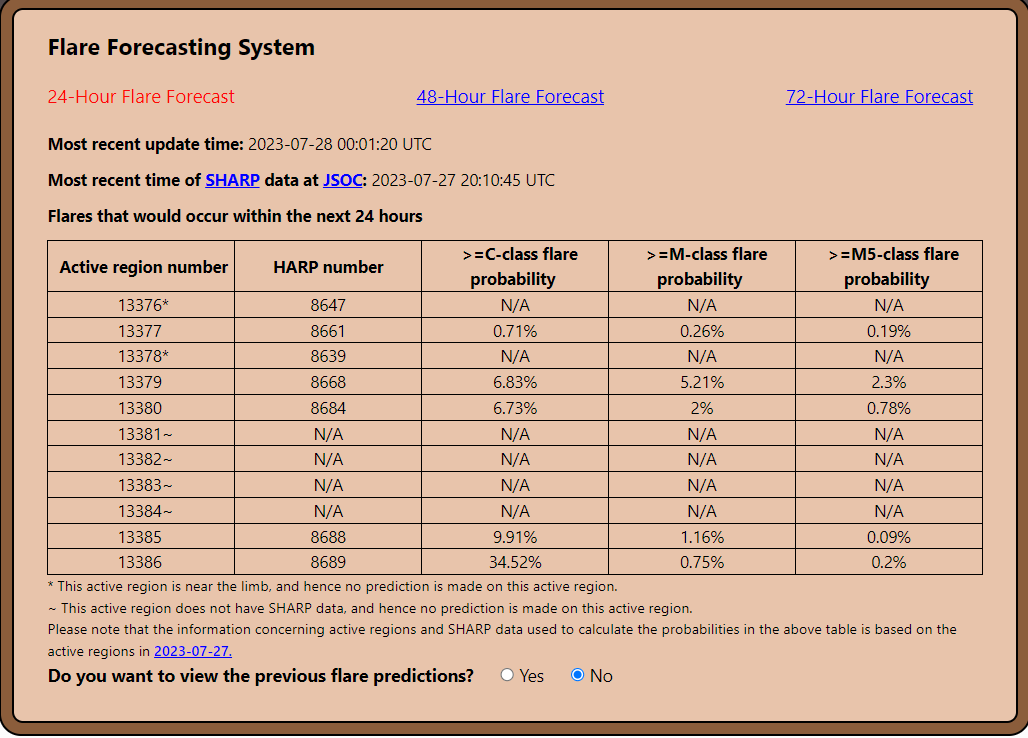}
	\caption{The graphical user interface of the SolarFlareNet system.}
	\label{fig:solarflarenetsystem}
\end{figure*}

We have implemented the probabilistic forecasting method described above into
an operational, near real-time flare forecasting system.
To access the system, visit the SolarDB website at \url{https://nature.njit.edu/solardb/index.html}. 
On the website, select and click the menu entry ``Tools'' and then select and click ``Flare Forecasting System.'' 
Figure~\ref{fig:solarflarenetsystem} shows the graphical
user interface (GUI) of the system.
It displays a probabilistic estimate of how likely an AR will produce a flare within the next 24, 48, and 72 hours of
the time point at which the system is updated each day.
No prediction is made for an AR marked with a special character *, \#, or $\sim$ where
\begin{itemize}
    \item * means the AR is near the limb,
\item \# means the AR is spotless with the number of spots being zero, 
\item $\sim$ means no SHARP data is available for the AR. 
\end{itemize}
The system provides daily predictions based on the data obtained from the previous day.
When the user clicks the link to the previous day, the user is led to the SolarMonitor site that is
accessible at \url{https://www.solarmonitor.org/index.php} where detailed AR information for that day is available. 
The system also provides previous forecasting results
since the operational system came online.
We compare the previous forecasting results with the true
flare events in the GOES X-ray flare catalogs provided by NCEI. 
The SolarFlareNet system achieves 89\% (76\%, 71\%, respectively) accuracy
for 24-hour (48-hour, 72-hour, respectively) ahead predictions.

\section{Discussion and Conclusion}

In this paper, we present a novel transformer-based framework
to predict whether a solar active region (AR) would produce a $\gamma$-class flare within the next 24 to 72 hours where $\gamma$ is $\geq$M5.0, $\geq$M, or $\geq$C. 
We use three transform models to handle the three classes of flares
individually and separately.
All three transformer models
perform binary predictions.
We collect ARs with flares that occurred between 2010 and 2022
from the GOES X-ray flare catalogs provided by
the National Centers for Environmental Information (NCEI).
In addition, we downloaded SHARP magnetic parameters from the Joint Science Operations Center (JSOC).
Each data sample contains SHARP parameters suitable for machine learning.
We conducted experiments using 10-fold cross-validation
\cite{Liu_2019FlarePrediction}.
Based on the experiments,
our transformer-based framework
generally performs better than
closely related methods
in terms of TSS (true skill statistics),
as shown in Table~\ref{tab:solarflarenetcomparativestudy2019sun3d}.
We further extend our framework to produce probabilistic forecasts of flares and
implement the framework into an
operational, near real-time flare forecasting system
accessible on the Web.
The probabilistic framework is comparable to a closely related method
\cite{Liu_2019FlarePrediction}
in terms of BS (Brier score) and
BSS (Brier skill score) when making 24-hour forecasts,
as shown in Table \ref{tab:probcalib},
although the existing method did not make 48- or 72-hour forecasts.
Thus, we conclude that SolarFlareNet is a feasible tool
for producing flare forecasts within 24 to 72 hours.

\section{Methods}

\subsection{Data Collection}\label{sec:SolarFlareNetdata}

\begin{table*}
\centering
\begin{tabular}{lll}
\hline
Keyword& Description & Formula \\ \hline
TOTUSJH & Total unsigned current helicity &  $H_{c_{total}} \propto \sum \vert B_z \cdot J_z \vert$ \\
TOTUSJZ & Total unsigned vertical current &   $J_{z_{total}} = \sum \vert J_z \vert dA$\\
USFLUX & Total unsigned flux &  $\Phi = \sum \vert B_z \vert dA$ \\
MEANALP & Mean characteristic twist parameter, $\alpha$ & $\alpha_{total} \propto \frac{\sum J_zB_z}{\sum B_z^2}$  \\
R\_VALUE & Sum of flux near polarity inversion line &  $\Phi=\sum \vert B_{LoS} \vert dA ~\textrm{within} ~R ~\textrm{mask}$ \\
TOTPOT & Total photospheric magnetic free energy density &  $\rho_{tot} \propto \sum (\pmb{B}^{\textrm{Obs}}-\pmb{B}^{\textrm{Pot}})^2dA$\\
SAVNCPP & Sum of the modulus of the net current per polarity &  $J_{z_{sum}} \propto \vert \sum^{B_z^+}J_zdA \vert + \vert \sum^{B_z^-}J_zdA \vert$ \\
AREA\_ACR & Area of strong field pixels in the active region &  $\textrm{Area} = \sum \textrm{Pixels} $\\
ABSNJZH & Absolute value of the net current helicity & $H_{c_{abs}} \propto \vert \sum B_z \cdot J_z \vert$  \\
\hline
\end{tabular}
\caption{Overview of the nine SHARP parameters used in our study.}
\label{tab:features_formulas}
\end{table*}

In this study we used SHARP magnetic parameters \cite{Bobra:2014SoPh..289.3549B,BobraCME2016ApJ...821..127B,Liu_2019FlarePrediction} 
downloaded from the Joint Science Operations Center (JSOC)  accessible at \url{http://jsoc.stanford.edu/}.
Specifically, we collect data samples, composed of SHARP parameters,
at a cadence of 12 minutes where
the data samples are retrieved from the {\sf hmi.sharp\_cea\_720s}
data series on the JSOC website
using the Python package SunPy \cite{SunPy2015CSD....8a4009S}.
We selected nine SHARP magnetic parameters as suggested in the literature \cite{Bobora:2015ApJ...798..135B,BobraCME2016ApJ...821..127B,Liu..Wang..Solar..2017ApJ...843..104L,Liu_2019FlarePrediction,Liu_2020CMEPrediction}.
These nine parameters include the
total unsigned current helicity (TOTUSJH), 
total unsigned vertical current (TOTUSJZ),
total unsigned flux (USFLUX),
mean characteristic twist parameter (MEANALP), 
sum of flux near polarity inversion line (R\_VALUE), 
total photospheric magnetic free energy density (TOTPOT), 
sum of the modulus of the net current per polarity (SAVNCPP), 
area of strong field pixels in the active region (AREA\_ACR), and
absolute value of the net current helicity (ABSNJZH).
Table \ref{tab:features_formulas} presents an overview of the
nine parameters.
The SHARP parameters' values are in different scales and units; therefore, we normalize each parameter's values 
using the min-max normalization method \cite{Liu_2020CMEPrediction,AbduallahSEP2022}. 
Formally, let $p^k_{i}$ be the original value of the $i$th parameter of the $k$th data sample.
Let $q^k_{i}$ be the normalized value of the $i$th parameter of the $k$th data sample.
Let $min_i$ be the minimum value of the $i$th parameter.
Let $max_i$ be the maximum value of the $i$th parameter.
Then
\begin{equation}
q^k_{i} = \frac{p^k_i - min_i}{max_i - min_i}
\end{equation}

We collected A-, B-, C-, M- and X-class flares that occurred between May 2010 and December 2022, 
and their associated active regions (ARs) from the GOES X-ray flare catalogs provided by
the National Centers for Environmental Information (NCEI). 
Flares without identified ARs were excluded.
This process yielded a database of
8 A-class flares,
6,571 B-class flares,
8,973 C-class flares,
895 M-class flares, and
58 X-class flares. 
Also, we collected 10 nonflaring ARs \cite{HSC-2020}.
We collected data samples that were 24 (48, 72, respectively) hours before a flare.
Furthermore, we collected data samples
that were 24 (48, 72, respectively) hours after the start time of each nonflaring AR.
The data was then cleaned as follows
\cite{BobraCME2016ApJ...821..127B,Liu_2019FlarePrediction,AbduallahSEP2022}. 

We discard ARs that are outside $\pm$ 70$^\circ$ of the central meridian. 
These ARs are near the limb and have projection effects
that render the calculation of the ARs' SHARP parameters incorrect.
In addition, we discard a data sample if
(i) its corresponding flare record has an absolute value of the radial velocity of SDO greater than 3500 m $s^{-1}$, 
(ii) the HMI data have low quality \cite{Hoeksema2014SoPh..289.3483HHMIQuality},
or (iii) the data sample has missing values or incomplete SHARP parameters. 
Thus, we exclude low-quality data
samples and keep qualified data samples of high
quality in our study.

\subsection{Data Labeling}\label{sec:SolarFlareNetdatalabeling}
\begin{figure*}
     \centering
     \includegraphics[width=0.49\linewidth]{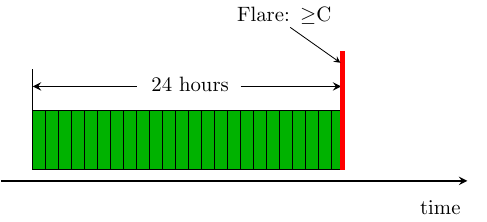}
     \includegraphics[width=0.49\linewidth]{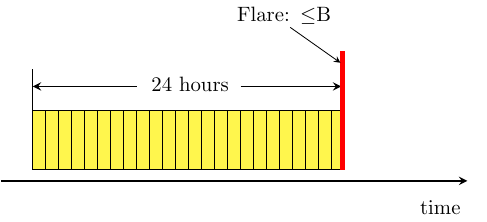}
         \caption{Illustration of positive and negative data samples used in predicting $\geq$C-class flares. In the left panel, the red vertical line indicates the start time of a $\geq$C-class flare. The data samples collected in the 24 hours prior to the red vertical line are labeled positive (in green color). In the right panel, the red vertical line
         indicates the start time of an A-class or B-class flare. The data samples collected in the 24 hours prior to the red vertical line are labeled negative
         (in yellow color).}
     \hfill
\label{fig:SolarFlareNetdatasamplespositivenegative}
\end{figure*}

Data labeling is crucial in machine learning.
To predict $\geq$C-class flares, suppose that a C-, M-, or X-class flare occurs at time point $t$ on an AR
(more precisely, the start time of the flare is $t$).
Data samples between $t$ and $t$ $-$ 24 hours (48, 72 hours, respectively) in the AR are labeled positive.
If the flare occurs at time point $t$ is an A-class or B-class flare,
the data samples between $t$ and $t$ $-$ 24 hours (48, 72 hours, respectively) in the AR are labeled negative.
Figure \ref{fig:SolarFlareNetdatasamplespositivenegative} illustrates the labeling scheme
to predict whether a $\geq$C-class flare would occur
within 24 hours.
In predicting $\geq$M-class flares,
we use $\geq$M-class flares to label positive data samples;
use $\leq$C-class flares to label negative data samples.
In predicting $\geq$M5.0-class flares,
we use $\geq$M5.0-class flares to label positive data samples;
use $\leq$C-class flares as well as M1.0- through M4.0-class flares to label negative data samples.
If there are recurring flares whose corresponding data samples overlap, 
we give priority to the largest flare and label the overlapped data samples based on the largest flare.
In all three prediction tasks, the data samples in the nonflaring ARs are labeled negative.

Table \ref{tab:positive_negative_counts} shows the total numbers of 
positive and negative data samples in each class for 24-,
48-, and 72-hour ahead flare predictions.
The numbers in the table are lower than expected.
This is because we discarded/removed
many low-quality data samples as described above. 
If a gap occurs in the middle of a time series due to removal, we
use a zero-padding strategy \cite{Liu_2019FlarePrediction,AbduallahSEP2022}
to create a
synthetic data sample to fill the gap. The synthetic data sample has
zero values for all nine SHARP parameters. The synthetic data
sample is added after normalization of the values of the SHARP parameters, and therefore the synthetic data sample does not affect the
normalization procedure.

For each prediction task, we divide the corresponding data samples into 10 equal sized distinct partitions/folds 
that are used to perform 10-fold cross-validation experiments.
In the run $i$, where $1 \leq i \leq 10$, 
we use fold $i$ as the test set and
the union of the other nine folds as the training set.
The data samples of the same AR are placed in the training set or the test set, but not both.
This scheme ensures that a model is trained with data different from the test data
and makes predictions on the test data that it has never seen during training.
There are 10 folds and, consequently, 10 runs.
The means and standard deviations of
the performance metrics' values over the 10 runs
are calculated and recorded.

\subsection{Data Augmentation}
\label{sec:SolarFlareNetdataaugmentation}
The data sets used in this study
to predict flares of the $\geq$M- and $\geq$M5.0-class are imbalanced
as shown in Table \ref{tab:positive_negative_counts}
where negative data samples
are much more than positive data samples. 
Imbalanced data pose a challenge in model training
and often result in poor model performance.
One may use data augmentation to combat the imbalanced data.
Data augmentation is an important technique that enriches training data and increases the generalization of the model~\cite{TransformerDataAug2021Deng}. 
Here, we adopt the Gaussian white noise (GWN) data augmentation scheme
because it has shown a significant improvement in
model performance\cite{CNNDataAugWearableSensor,CNNGlucoseDataAug2020}. 
GWN assumes that any two values are statistically independent, regardless of how close they are in time. 
The stationary random values of GWN are generated using the zero mean and 5$\%$ of the standard deviation. 
During training, the data augmentation is applied to the minority (positive) class, 
leaving the majority (negative) class as is. 
During testing, the data are left without any augmentation
so that the model predicts only on the actual test data
to avoid any misleading performance assessment.

\begin{table*}
\centering
\begin{tabular}{ccccc}
\hline
     Hour     & Data samples& $\geq$M5.0 class  & $\geq$M class  & $\geq$C class  \\  
\hline
24 
   & Positive &  2,125 & 13,989 & 244,968  \\
   & Negative &  461,060 & 449,196 & 218,517 \\
   \hline
48  
   & Positive &  2,255 & 16,709 & 316,149  \\
   & Negative & 615,708 & 602,154 & 304,714  \\
   \hline
72 
   & Positive &  2,375 & 18,505 & 356,219  \\
   & Negative & 704,997 & 689,567 & 350,953 \\
   \hline
\end{tabular}
\caption{Total numbers of positive and negative data samples in each class for 24-,
48-, and 72-hour ahead flare predictions.}
\label{tab:positive_negative_counts}
\end{table*}

\subsection{The SolarFlareNet Architecture}\label{sec:SolarFlareNetarchitecture}

\begin{figure*}[th!]
    \centering
    \includegraphics[]{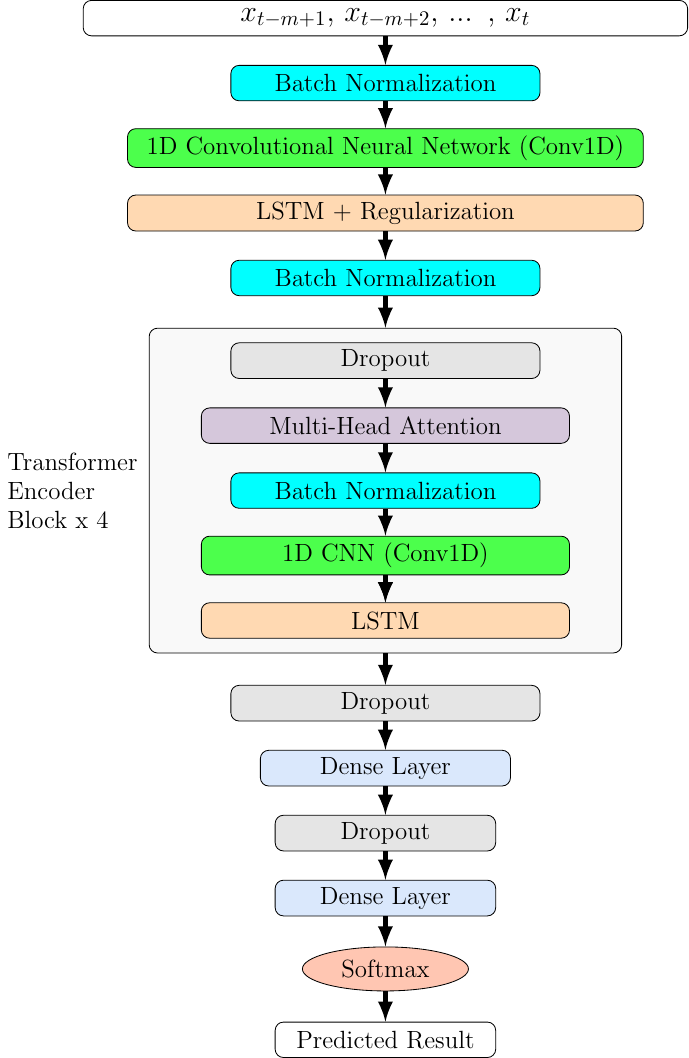}
     \caption{Architecture of SolarFlareNet.
    }
    \label{fig:SFNarchitecture}
\end{figure*}

Figure~\ref{fig:SFNarchitecture} presents the architecture of SolarFlareNet. 
It is a transformer-based framework that combines
a one-dimensional convolutional neural network (Conv1D), 
long short-term memory (LSTM), 
transformer encoder blocks (TEBs), 
and additional layers that include
batch normalization (BN) layers, dropout layers, and dense layers. 
The first layer is the input layer, which takes as input
a time series of $m$ consecutive data samples
$x_{t-m+1}$, $x_{t-m+2}$ \dots $x_{t-1}$, $x_{t}$ 
where $x_{t}$ is the data sample at time point $t$ \cite{AbduallahSEP2022}. 
(In the study presented here, $m$ is set to 10.)
The input layer is followed by a BN layer.  
BN is an additional
mechanism to stabilize SolarFlareNet, make it faster, 
and help to avoid overfitting during training \cite{TransoferEncoderOnly2021}. 
We applied BN after the input layer,
the LSTM layer,
and within the TEBs
to make sure that SolarFlareNet is stable
throughout the training process. 
The BN layer is followed by the Conv1D layer
because time series generally have a strong 1D time locality
that can be extracted by the Conv1D layers \cite{MosheCyberAttachsUsing1DCNN2018}.
Then, the LSTM layer is used, which is equipped with regularization to also avoid overfitting. 
LSTM is suitable for handling time series data to capture the temporal correlation and dependency in the data. 
Adding an LSTM layer after a Conv1D layer has shown significant improvement in
time series prediction\cite{TSINet_FLAIRS_Abduallah_Wang_Shen_Alobaid_Criscuoli_Wang_2021,AbduallahKpNetTransformer2022,Abduallah_Wang_Bose_Zhang_Gerges_Wang_2022_DST}. 
The LSTM layer passes the learned features and patterns to
a BN layer to stabilize the network before the data go to the TEBs.

We use transformer encoders without decoders
because we process time series here, rather than performing natural language processing
where the decoders are required to decode the words for sentence translation. 
The number of TEBs is set to 4.
This number has a significant effect
on the overall performance of the model \cite{AttentionAllYouNeedNIPS2017_3f5ee243}.
When we use less than 4 TEBs,
the model is not able to learn useful patterns
and is under-fitted.
When we use more than 4 TEBs,
the large number of TEBs causes overhead on the encoder processing while the model
tends to do excessive overfitting and lean toward
the majority class (i.e., negative class)
in the data, ignoring the minority class
(i.e., positive class) entirely. 
Each TEB is configured with a dropout layer, multi-head attention (MHA) layer, a BN layer, a Conv1D layer, and an LSTM layer. The MHA layer is the most important layer in the encoder because it provides the transformation on the sequence values to obtain the different metrics. The MHA layer is configured with 4 heads and each attention head is also set to 4. 
The dropout layer is mainly used to overcome the overfitting caused by the imbalanced data. 
It drops a percentage of the neurons from the architecture,
which causes the internal architecture of the model to change, 
allowing for better performance and stability. 
Finally, the softmax function is used as the final activation function, 
which produces a probabilistic estimate of how likely a flare
will occur within the next 24 (48, 72, respectively)
hours of $t$.

\subsection{Ablation Study}\label{sec:ablationstudy}

\begin{table*}
\centering
\begin{tabular}{cllccc}
\hline
     Hour     &  & Method & $\geq$M5.0 class  & $\geq$M class  & $\geq$C class  \\  
\hline
24&     & SolarFlareNet   & \textbf{0.818} (0.021) & \textbf{0.839} (0.030)  & \textbf{0.835} (0.048)\\
        &       & SolarFlareNet-Conv & 0.780 (0.036) &   0.759 (0.052) &   0.822 (0.023)         \\
        &       & SolarFlareNet-L       &  0.779 (0.022) &   0.737 (0.041) &   0.713 (0.039)  \\        
        
        &       & SolarFlareNet-ConvL       & 0.742 (0.029) &   0.719 (0.041) &   0.728 (0.037) \\
        &       & SolarFlareNet-T       &  0.716 (0.101) & 0.704 (0.093) & 0.712 (0.078) \\  
         \hline
         48 &  & SolarFlareNet & \textbf{0.736} (0.112) &\textbf{0.728} (0.090)  & \textbf{0.719} (0.079) \\
        &       & SolarFlareNet-Conv   &0.729 (0.049)&   0.715 (0.055) &   0.695 (0.035)        \\
        &       & SolarFlareNet-L       & 0.694 (0.066) &   0.689 (0.012) &   0.675 (0.021)  \\        
        
        &       & SolarFlareNet-ConvL       & 0.681 (0.063) &   0.676 (0.054) &   0.673 (0.048)\\
        &       & SolarFlareNet-T       & 0.662 (0.061) &   0.647 (0.032) &   0.641 (0.033)  \\  
         \hline
        72&     & SolarFlareNet   & \textbf{0.729} (0.108) & \textbf{0.714} (0.095)& \textbf{0.709} (0.058)  \\
        &       & SolarFlareNet-Conv    &  0.703 (0.042) & 0.696 (0.011) & 0.658 (0.023)     \\
        &       & SolarFlareNet-L        &  0.688 (0.046)  & 0.666 (0.039) & 0.658 (0.016) \\        
        
        &       & SolarFlareNet-ConvL       & 0.665 (0.026)& 0.643 (0.031) & 0.632 (0.030)\\
        &       & SolarFlareNet-T       &  0.635 (0.028)& 0.624 (0.046) & 0.619 (0.033)  \\        
         \hline
\end{tabular}
\caption{TSS values of the five methods considered in the ablation study.}
\label{tab:ablationtestsresults}
\end{table*}

We performed ablation tests to assess each component of SolarFlareNet.
We consider four variants of SolarFlareNet, denoted
SolarFlareNet-Conv, 
SolarFlareNet-L, 
SolarFlareNet-ConvL, 
and SolarFlareNet-T, 
respectively.
Here,
SolarFlareNet-Conv 
(SolarFlareNet-L, 
SolarFlareNet-ConvL, 
SolarFlareNet-T, respectively)
represents the subnet of SolarFlareNet in which the Conv1D layer
(LSTM layer, 
Conv1D and LSTM layers, 
transformer network with the 4 TEBs,
respectively) 
is removed while keeping the remaining components of the SolarFlareNet framework.
Table \ref{tab:ablationtestsresults} 
compares the TSS values of the five models
for the 24-, 48-, and 72-hour ahead flare prediction. 
It can be seen from Table \ref{tab:ablationtestsresults} that the full model, SolarFlareNet, 
 outperforms the four subnets in terms of the TSS metric. 
 This happens because the SolarFlareNet-Conv model captures the temporal correlation of the test data, but does not learn additional characteristics of the data to build a stronger relationship between the test data. 
 SolarFlareNet-L captures the properties of the test data, but lacks knowledge of the temporal correlation patterns in the data to deeply analyze the sequential information in the test data. 
 It can also be seen from Table \ref{tab:ablationtestsresults} that the SolarFlareNet-ConvL model is not as good as the full model, 
 indicating that the transformer network alone is not enough to produce the best results. Lastly, SolarFlareNet-T has the least performance among the four subnets, demonstrating the importance of the transformer network. 
 In conclusion, our ablation study indicates that the performance of the proposed SolarFlareNet framework is not dominated by any single component.
 In fact, all components have made contributions to the overall performance of the proposed framework.

\bibliographystyle{icml2020}


\end{document}